\documentclass[numberedappendix, preprintnumbers, amsmath,amssymb,nofootinbib]{emulateapj} 

\usepackage{graphicx}
\usepackage{dcolumn}
\usepackage{bm}
\usepackage{amsmath,amsfonts,amssymb,slashed,upgreek,color}
\usepackage{subfigure}
\usepackage{IEEEtrantools}
\usepackage{color}
\usepackage{float}

\newcommand\bie{\begin{IEEEeqnarray}{rCl}}
\newcommand\eie{\end{IEEEeqnarray}}
\newcommand\ee{\end{equation}}
\newcommand\be{\begin{equation}}
\newcommand\eea{\end{eqnarray}}
\newcommand\bea{\begin{eqnarray}}

\newcommand\lsim{\mathrel{\rlap{\lower4pt\hbox{\hskip1pt$\sim$}}
        \raise1pt\hbox{$<$}}}
\newcommand\gsim{\mathrel{\rlap{\lower4pt\hbox{\hskip1pt$\sim$}}
        \raise1pt\hbox{$>$}}}

\newcommand{\bq}{\mathbf{q}}
\newcommand{\bk}{\mathbf{k}}

\newcommand{\eps}{\epsilon}
\newcommand{\bs}{\mathbf{s}}
\newcommand{\br}{\mathbf{r}}
\newcommand{\bp}{\mathbf{p}}
\newcommand{\PP}{{\cal{P}}}
\newcommand{\bku}{\mathbf{k}\in {\rm uhs}}
\newcommand{\bkku}{{\mathbf{p},\mathbf{q}\in {\rm uhs}}}

\usepackage{color}
\usepackage{soul}
\usepackage[makeroom]{cancel}
\definecolor{darkred}{rgb}{0.9,0.1,0.1}
\definecolor{darkgreen}{rgb}{0.0,0.7,0.0}
\definecolor{darkblue}{rgb}{0.0,0.0,1.0}

\usepackage{hyperref}

\shorttitle{Three-point phase correlations} 

\shortauthors{Wolstenhulme et al.}

\begin{document}

\title{Three-point phase correlations: A new measure of non-linear large-scale structure}

\author{Richard Wolstenhulme$^{1}$, Camille Bonvin$^{1,2,3}$ and Danail Obreschkow$^{4}$\\  }
\affiliation{${}^{1}$ Kavli Institute for Cosmology Cambridge and Institute of Astronomy,
Madingley Road, Cambridge CB3 OHA, United Kingdom\\
${}^{2}$ DAMTP, Centre for Mathematical Sciences, Wilberforce Road, Cambridge CB3 OWA, United Kingdom\\
${}^{3}$ CERN, Theory Division,  1211 Geneva, Switzerland\\
${}^{4}$International Centre for Radio Astronomy Research (ICRAR), M468, University of Western Australia, 35 Stirling Hwy, Crawley, WA 6009, Australia}

\date{\today}  

\begin{abstract}
    We derive an analytical expression for a novel large-scale structure observable: the line correlation function. The line correlation function, which is constructed from the three-point correlation function of the phase of the density field, is a robust statistical measure allowing the extraction of information in the non-linear and non-Gaussian regime. We show that, in perturbation theory, the line correlation is sensitive to the coupling kernel $F_2$, which governs the non-linear gravitational evolution of the density field. We compare our analytical expression with results from numerical simulations and find a 1-$\sigma$ agreement for separations $r\gsim 30\, h^{-1} {\rm Mpc}$. Fitting formulae for the power spectrum and the non-linear coupling kernel at small scales allow us to extend our prediction into the strongly non-linear regime where we find a 1-$\sigma$ agreement with the simulations for $r \gsim 2 h^{-1}\mathrm{Mpc}$. We discuss the advantages of the line correlation relative to standard statistical measures like the bispectrum. Unlike the latter, the line correlation is independent of the bias, in the regime where the bias is local and linear. Furthermore, the variance of the line correlation is independent of the Gaussian variance on the modulus of the density field. This suggests that the line correlation can probe more precisely the non-linear regime of gravity, with less contamination from the power spectrum variance.
\end{abstract}


\maketitle


\section{Introduction}
\label{sec:intro}

Galaxy redshift surveys provide fundamental information about the constituents of our universe (dark matter and dark energy), and about the gravitational laws that govern its evolution. Fluctuations in the galaxy number count for example can be used to measure the growth rate of large-scale structure and to test the consistency of the standard $\Lambda$CDM cosmological model (see e.g. the latest results from BOSS~\citep{Samushia2014}). 

The crucial step in extracting cosmological parameters from three-dimensional galaxy maps is the use of statistical measures. These measures filter out the random component of the galaxy distribution, allowing us to access the cosmological information imprinted in the maps. The simplest and most extensively used statistical measure is the two-point correlation function of the galaxy density field; or equivalently its Fourier transform, the power spectrum. 

If the universe is statistically homogeneous and isotropic and if the galaxy density field is Gaussian, the power spectrum measures all of the cosmological information contained in three-dimensional galaxy maps. Consequently at sufficiently large scales, in the regime where the density field remains linear up to today, the power spectrum is sufficient to completely characterize matter fluctuations in our universe~\footnote{This assumes that primordial perturbations are Gaussian.}. At small scales however, the situation is more complicated: the non-linear gravitational evolution of density perturbations generates non-Gaussianity in the density field and therefore the power spectrum does not measure all of the available information~\citep[e.g.][]{2013MNRAS.429..344K}. Additional statistical measures are necessary to fully describe the density field. Various statistics have been proposed, such as higher-order correlation functions~\citep[e.g.][]{Frieman:1993nc, 1994PhRvL..73..215F, Frieman:1999qj, Szapudi:2001mh, Takada:2002qq, Gaztanaga:2005an}: in particular the bispectrum~\citep[e.g.][]{Scoccimarro1998a, Scoccimarro:2000sp, Verde:2001sf, Smith:2005ap, Gil-Marin2014}, Minkowski functionals~\citep{Mecke1994a, Gott1987a, Weinberg1987a, Melott1988a, 1989ApJ...340..625G, Hikage2003a, Codis2013}, genus statistics~\citep{1986ApJ...306..341G, Hoyle:2002jm, Hikage:2002ki, Park:2005fk}, void distribution function~\citep{1979MNRAS.186..145W}, fractal correlation dimension~\citep{Scrimgeour:2012wt}, etc. The motivation to find simple and powerful statistical measures beyond the power spectrum is strong, because a major amount of information escapes from the two-point statistics during the non-linear gravitational evolution \citep{Carron2012}.

In this paper we focus on a new statistical measure: the line correlation function, which we recently presented in~\citet{Obreschkow2013a}. The line correlation function is constructed from the three-point correlation of the phase of the density field. If the density field is Gaussian, the phases do not contain any information: they are by definition random and uncorrelated. All of the information is in the amplitude of the density field (which can be measured with the power spectrum) and the line correlation simply vanishes. If the density field is non-Gaussian however, this is not the case anymore: correlations between the phases emerge, encapsulating information beyond the power spectrum. 

Phase information has been studied by a number of authors and various ways of measuring it have been proposed. \citet{1991ApJ...377...29S} used the distribution of phase maxima to differentiate Gaussian from non-Gaussain density fields. In~\citet{Ryden:1991fi},~\citet{Soda:1991jz} and~\citet{Jain:1998it} the non-linear gravitational evolution of the phases and the shift it induces on individual modes have been calculated. \citet{Chiang:1999hm} explored the use of the phase difference between Fourier modes as a measure of phase information. This idea has then been investigated in a number of subsequent papers~\citep[e.g.][]{Chiang:2002sb, Chiang:2002vm, Watts:2002py, Chiang:2000vt, Coles:2000zg}. \citet{Matsubara2003c} proposed to use the probability distribution function of the phase sum as a new statistical measure of phase information. He expressed the probability distribution function in terms of polyspectra, using perturbation theory (hereafter PT). This new measure was then explored in numerical simulations~\citep{Hikage2004} and applied for the first time to SDSS galaxies~\citep{Hikage2005a}. In addition to these works related to galaxy clustering, \citet{Coles:2003dw} proposed to use the phases of the temperature fluctuations in the Cosmic Microwave Background as a probe of non-Gaussian initial conditions. \citet{Szepietowski2014} on the other hand studied the use of the phases to improve the reconstruction of the density field from lensing maps.
These works clearly demonstrate the potential of phase correlations to probe information beyond the linear, Gaussian regime. 

Our paper builds on these previous results and presents an analytical description of the line correlation function. As explained in~\citet{Obreschkow2013a}, the line correlation function provides a new way of measuring phase correlations. It is defined as the three-point correlation function (in real space) of the phases of the density field, equidistantly distributed on a line: 
\be
\ell(r)\propto \big\langle \eps(\bs)\eps(\bs+\br)\eps(\bs-\br) \big\rangle\, ,
\ee
where $\eps(\bk)=\delta(\bk)/|\delta(\bk)|$ is the phase of the density field $\delta$, and $\eps(\br)$ is its inverse Fourier transform. Here $\langle\, \rangle$ denotes a statistical or ensemble average over all possible realisations. This average separates the useful information present in the phases (the patterns created by non-linear gravitational evolution) from the background noise coming from uncorrelated random phases. Note that the translation vector $\bs\in\mathbb{R}^3$ and the direction of the vector $\br$ can be chosen arbitrarily by virtue of the assumed statistical homogeneity and isotropy. The crucial advantage of the line correlation with respect to the bispectrum is that, by construction, it is independent of the modulus of the density field $|\delta(\bk)|$. It provides therefore a more direct way to target information beyond the power spectrum. 

In~\citet{Obreschkow2013a}, we showed, based on a set of cosmological simulations, that the line correlation function can be used to measure the temperature of dark matter and to identify filamentary structures in our universe. Here, our goal is to find an analytical expression for $\ell(r)$ using PT. Such an expression is extremely useful to assess how sensitive the line correlation is to cosmological parameters, and which kind of degeneracies we can expect to break with this new estimator. 

The rest of the paper is organized as follows. In section~\ref{sec:distribution} we briefly present the calculation of the probability distribution function of the phases of the density field, following the work of~\citet{Matsubara2003c}. In section~\ref{sec:3points}, we use this probability distribution function to calculate the ensemble average of the three-point correlation function of the phase in PT. In section~\ref{sec:line} we calculate the line correlation function and in section~\ref{sec:num} we compare our result with numerical simulations. We conclude and discuss future directions in section~\ref{sec:conclusion}.

\section{The distribution function of the phases}
\label{sec:distribution}

Our observable is the line correlation function $\ell(r)$, defined as (for more detail see~\citet{Obreschkow2013a})
\begin{align}
\label{ldef}
\ell(r)=&\frac{V^3}{(2\pi)^9}\left(\frac{r^3}{V}\right)^{3/2}\big\langle \eps(\bs)\eps(\bs+\br)\eps(\bs-\br) \big\rangle\\
=&\frac{V^3}{(2\pi)^9}\left(\frac{r^3}{V}\right)^{3/2}\hspace{-0.3cm}\underset{|\bk_1|, |\bk_2|, |\bk_3|\leq 2\pi/r}{\int\!\!\int\!\!\int\!\!} 
\!\!\!\!\!\!d^3\bk_1\,d^3\bk_2\,d^3\bk_3\nonumber\\
&e^{i\big[\bk_1\cdot \bs+\bk_2\cdot(\bs+\br)+\bk_3\cdot(\bs-\br) \big]} \big\langle \eps(\bk_1)\eps(\bk_2)\eps(\bk_3) \big\rangle\, .\nonumber
\end{align}
Here $\eps(\br)$ is the inverse Fourier transform of the phase $\eps(\bk)=\delta(\bk)/|\delta(\bk)|$, which is in our convention
\be
\eps(\br)=\int d^3\bk\, e^{i\bk\cdot\br}\eps(\bk)\, ,
\ee
and $V$ is the survey volume~\footnote{We will see that the line correlation function as defined in eq.~\eqref{ldef} is actually independent of the survey volume $V$. Note that we use here a different Fourier convention than in~\citet{Obreschkow2013a} resulting in an additional factor $V^3/(2\pi)^9$ in eq.~\eqref{ldef}.}. As discussed in~\citet{Obreschkow2013a}, we must introduce a cut-off at high frequency (here $|\bk_i|\leq 2\pi/r$) to avoid divergence of the line correlation by virtue of an infinite number of phase factors at arbitrarily large $k$-vectors. Note that our choice of cut-off differs slightly from the one used in~\citet{Obreschkow2013a} (where $|\bk_1|, |\bk_2|\leq \pi/r$ and $|\bk_1+\bk_2|\leq 2\pi/r$). Here all modes are smaller than $2\pi/r$ so that the result is invariant over permutations of $\bk_1, \bk_2$ and $\bk_3$.  In addition, we regularize the situation at large scales (where the phases are also randomly distributed), by introducing the prefactor  $(r^3/V)^{3/2}$ in eq.~\eqref{ldef}.

The line correlation function $\ell(r)$ depends directly on the three-point correlation function of the phase $ \langle \eps(\bk_1)\eps(\bk_2)\eps(\bk_3) \rangle$. 
The challenge in evaluating this expression is the difficulty of calculating the expectation of a ratio of statistical fields. In particular
\bea
\label{ratio}
\langle\eps(\bk_1)\eps(\bk_2)\eps(\bk_3) \rangle&=&
\left\langle\frac{\delta(\bk_1)\delta(\bk_2)\delta(\bk_3)}{|\delta(\bk_1)\delta(\bk_2)\delta(\bk_3)|} \right\rangle\\
&\neq& \frac{\langle\delta(\bk_1)\delta(\bk_2)\delta(\bk_3) \rangle}{\langle|\delta(\bk_1)\delta(\bk_2)\delta(\bk_3)| \rangle}\, .\nonumber
\eea
We address this challenge by expressing the three-point phase correlation in terms of the probability distribution function of the phase $\PP[\theta]$ (here $\PP$ is a functional of the field $\theta(\bk)$ defined through $\eps(\bk)=e^{i\theta(\bk)}$):
\be
\label{3pointphase}
\langle \eps(\bk_1)\eps(\bk_2)\eps(\bk_3) \rangle=\int [d\theta]\PP[\theta]\eps(\bk_1)\eps(\bk_2)\eps(\bk_3)\, .
\ee
To calculate the three-point phase correlation we need therefore an expression for the probability distribution function $\PP[\theta]$. In a Gaussian density field, the phases are uniformly distributed between 0 and $2\pi$ so that $\PP[\theta][d\theta]=[d\theta]/2\pi$. The integral of each of the modes in eq.~\eqref{3pointphase} gives
\be
\int_0^{2\pi}\frac{d\theta(\bk_i)}{2\pi}\eps(\bk_i)=0,\quad i=1,2,3\, ,
\ee
and the three-point correlation function vanishes~\footnote{Note that if $\bk_1=-\bk_2$ and $\bk_3=0$, $\langle \eps(\bk_1)\eps(\bk_2)\eps(\bk_3) \rangle=1$. This is however not an interesting case since it corresponds to an homogeneous mode $\bk_3=0$, which is part of the background.}. On the other hand, for a non-Gaussian density field, the distribution function is non-uniform and this generates correlations between the phases in eq.~\eqref{3pointphase}. 

The probability distribution function for a mildly non-Gaussian field, has been calculated in~\citet{Matsubara2003c}. Here we recall the main steps of the derivation.
We start with the probability distribution function for the density field $\delta(\bk)$, which reads~\citep{Matsubara1995a}
\bie
\label{prob_gen}
\mathcal{P}[&\delta&] = \exp \Bigg[ \sum_{N=3}^{\infty}\frac{(-1)^N}{N!} \int d^3\bk_1 \dots d^3\bk_N \\
&&  \times \langle \delta(\bk_1) \dots  \delta(\bk_N) \rangle_c \frac{\partial}{\partial \delta(\bk_1)}\dots \frac{\partial}{\partial \delta(\bk_N)}  \Bigg]\mathcal{P}_G[\delta]\, , \nonumber
\eie
where $\langle \delta(\bk_1) \dots  \delta(\bk_N) \rangle_c$ refers to the `connected' part of the $N$th moment and $\PP_G[\delta]$ is the Gaussian probability distribution function.

For a mildly non-Gaussian field, one can expand the exponential and keep only the first term, which is the dominant contribution. This gives rise to the so-called Edgeworth expansion~\citep{1991ApJ...381..349S, 1995ApJ...442...39J, 1995ApJ...443..479B}
\begin{align}
\label{P_Edg}
&\mathcal{P}[\delta] =N_G \exp \left(-\frac{1}{2}\int d^3\bk \frac{\delta(\bk) \delta(-\bk)}{P(k)} \right)\\
&\!\! \left\{1 + \frac{1}{3!}\! \int\!\! d^3\bp\, d^3\bq \, \frac{B(\bp,\bq, -\bp-\bq)\delta(-\bp)\delta(-\bq)\delta(\bp+\bq)}{P(p)P(q)P(|\bp+\bq|)} \right\} \nonumber
\end{align}
where $N_G$ is a normalisation factor. Here $P(k)$ and $B(\bp,\bq,\bk)$ are the power spectrum and bispectrum defined through
\bea
\langle \delta(\bk)\delta(\bk')\rangle&=&P(k)\delta_D(\bk+\bk')\, ,\\
\langle \delta(\bp)\delta(\bq)\delta(\bk)\rangle&=&B(\bp,\bq,\bk)\delta_D(\bp+\bq+\bk)\, .
\eea
Note that by statistical homogeneity and isotropy the bispectrum depends only on three independent variables characterising the triangle, e.g. three lengths ($p, q$ and $|\bp+\bq|$), or two lengths and one angle ($p, q$ and $\hat\bp\cdot\hat\bq$). Expression~\eqref{P_Edg} is valid in the mildly non-linear regime and is expected to break down at very non-linear scales.

We can now discretize the density field for a finite survey volume $V$: $\delta(\bk)\rightarrow \delta_\bk$ and the integrals becomes $\int d^3\bk \rightarrow \frac{(2\pi)^3}{V} \sum_{\bk}$. The distribution function becomes a multivariate distribution for the discrete set $\{\delta_\bk\}$. Since the density field $\delta(\br)$ is real, the variables $\delta_\bk$ and $\delta_{-\bk}$ are not independent; in fact, $\delta_{-\bk}=\delta^*_\bk$. We can therefore restrict $\bk$ to the upper hemisphere (uhs $~\equiv\{k_z>0\}\cup\{k_z=0,~k_y>0\}\cup\{k_z=k_y=0,~k_x>0\}$), where ${\rm Re}(\delta_\bk)$ and ${\rm Im}(\delta_\bk)$ are independent and the degrees of freedom in the lower hemisphere (lhs) are simply relabelled using the reality condition $\delta_{\bk} = \delta^*_{-\bk}$. The probability distribution function becomes
\begin{align}
\mathcal{P}&(\{\delta_\bk,\delta^*_\bk \})\prod_{\bku} d\delta_\bk d\delta^*_\bk=\\
& N_G \prod_{\bku} d\delta_\bk d\delta^*_\bk \exp\left[-\frac{(2\pi)^3}{V}\frac{\delta_\bk \delta^*_{\bk}}{P(k)}\right]\nonumber\\
&\Bigg\{1 + \frac{(2\pi)^6}{3 V^2} 
\sum_\bkku\Bigg[ \frac{B\big(p,q, |\bp+\bq|\big)\delta_{-\bp}\delta_{-\bq}\delta_{\bp+\bq}}{P(p)P(q)P(|\bp+\bq|)}\nonumber\\
&+\frac{B\big(p,q, |\bp-\bq|\big)\delta_{-\bp}\delta_{\bq}\delta_{\bp-\bq}}{P(p)P(q)P(|\bp-\bq|)} \Bigg]\Bigg\}\nonumber\, .
\end{align}
We then do a change of variables from $\{\delta_\bk, \delta^*_\bk \}$ to $\{|\delta_\bk|, \theta_\bk \}$ and we integrate over the moduli $|\delta_\bk|$. Using that the distribution function of the phase has to be real we obtain
\begin{align}
&\mathcal{P}(\{\theta_\bk \})\!\!\!\prod_{\bku} \!\!\!d\theta_\bk=\!\Bigg\{1+\frac{\sqrt{\pi}}{6}\!\!\!\sum_{\bp\in{\rm uhs}}\!\!
b(p,p, 2p)\cos(2\theta_{\bp}\!-\!\!\theta_{2\bp}\!)\nonumber\\
&+\frac{1}{3}\left(\frac{\sqrt{\pi}}{2} \right)^3\!\!\!\sum_{\bp\neq \bq\in {\rm uhs}}
 \!\!\!\!\Big[b\big(p,q,|\bp+\bq|\big)\cos(\theta_{\bp}+\theta_{\bq}-\theta_{\bp+\bq})\nonumber\\
&+ b\big(p,q,|\bp-\bq|\big)\cos(\theta_{\bp}-\theta_{\bq}+\theta_{\bp-\bq})\Big] \Bigg\}
\prod_{\bku} \frac{d\theta_\bk}{2\pi}\, , \label{Probability}
\end{align}
where we have defined
\be
\label{defb}
b\big(p,q,k)\equiv\sqrt{\frac{(2\pi)^3}{V}}\frac{B(p,q,k)}{\sqrt{P(p)P(q)P(k)}}\, .
\ee

\section{The phase correlation function}
\label{sec:3points}

We can now use the probability distribution function~\eqref{Probability} to calculate the correlation functions of the phase. 

\subsection{The one-point correlation function}

Let us start by looking at the mean of the phase $\langle \eps_{\bk_1} \rangle$ for a mode $\bk_{1}$ in the upper hemisphere
\be
\langle \eps_{\bk_1}\rangle=\prod_{\bku}\int d\theta_{\bk} \mathcal{P}(\{{\theta_{\bk}}\})  e^{i\theta_{\bk_1}}.
\ee
The Gaussian part of the probability distribution function gives
\be
\langle \eps_{\bk_1}\rangle_G=\int \frac{d\theta_{\bk_1}}{2\pi}e^{i\theta_{\bk_1}}\times\prod_{\substack{\bku\\  \bk\neq \bk_1}}\int \frac{d\theta_{\bk}}{2\pi}\,\,=0\, .
\ee
As expected, if the phases are randomly distributed between $0$ and $2\pi$, the average of $\eps=e^{i\theta}$ is zero.

To understand the contribution from the non-Gaussian part of the probability distribution function, let us first examine the first line of eq.~\eqref{Probability}. It gives a contribution of the form
\begin{align}
\sum_{\bp\in{\rm uhs}}\!b(p,p, 2p)\!\int \!\frac{d\theta_{\bk_1}}{2\pi}e^{i\theta_{\bk_1}}\!\times\!\prod_{\substack{\bku\\  \bk\neq \bk_1}}\int \frac{d\theta_{\bk}}{2\pi}\cos(2\theta_{\bp}-\theta_{2\bp})\nonumber\, .
\end{align}
The integral over $\theta_{\bk_1}$ vanishes for all values of $\bp$ except when $2\bp=\bk_1$. However in this case, in the product over $\bk\neq\bk_1$ there are two remaining integrals of the form
\begin{align}
&\int \frac{d\theta_{(\bk_1/2)}}{2\pi}\cos(2\theta_{(\bk_1/2)})=0\, ,\nonumber\\
&\int \frac{d\theta_{(\bk_1/2)}}{2\pi}\sin(2\theta_{(\bk_1/2)})=0\, .\nonumber
\end{align}
Hence the first line in eq.~\eqref{Probability} does not contribute to the mean of the phase $\langle \eps_{\bk_1} \rangle$. A similar argument shows that the second and third line in eq.~\eqref{Probability} do not contribute either~\footnote{Note that if $\bk_1=0$, the non-Gaussian part of the distribution function contributes to the mean. This mode is however part of the background and should not be treated as a perturbation.}. 

For a mode $\bk_1$ in the lower hemisphere, we can simply rewrite
\be
\langle \eps_{\bk_1}\rangle=\langle \eps^*_{-\bk_1}\rangle=\prod_{\bku}\int d\theta_{\bk} \mathcal{P}(\{{\theta_{\bk}}\})  e^{-i\theta_{-\bk_1}}\, ,
\ee
and we also find  
\be
\label{eps1}
\langle \eps_{\bk_1} \rangle=0\, .
\ee

This shows that even though the probability distribution function of the phase becomes non-uniform in the non-linear regime, the average of the phase factor over different realisations of the density field remains zero. Note that this is not in contradiction with previous works~\citep{Ryden:1991fi, Soda:1991jz, Jain:1998it}, which established that non-linearities shift the phases by a subsequent amount. Here we simply show that this shift of individual phases is such that it preserves their mean. 
It can be interpreted as a consequence of the statistical isotropy of the density field, which prevents the phase to acquire any preferred direction on average.

Therefore, we cannot probe the non-uniformity of the probability distribution function~\eqref{Probability}, by measuring the mean of the phase of a homogeneous and isotropic field. 

\subsection{The two-point correlation function}

We can then look at the two-point phase correlation $\langle \epsilon_{\bk_1}\epsilon_{\bk_2}\rangle$. 
We first consider the case where $\bk_1 \in \mathrm{uhs}$ and $\bk_2 \in {\rm lhs}$ (i.e. $-\bk_2\in {\rm uhs}$) 
\begin{equation}
\langle \epsilon_{\bk_1}\epsilon_{\bk_2}\rangle = \langle \epsilon_{\bk_1}\epsilon^*_{-\bk_2}\rangle
=\prod_{\bku}\int d\theta_{\bk} \mathcal{P}(\{{\theta_{\bk}}\})  e^{i\theta_{\bk_1}} e^{-i\theta_{-\bk_2}}. \end{equation}
The Gaussian part of the probability distribution function gives for $\bk_1\neq -\bk_2$
\begin{align}
\langle \epsilon_{\bk_1}\epsilon_{\bk_2}\rangle_G=&\int \frac{d\theta_{\bk_1}}{2\pi}e^{i\theta_{\bk_1}}\int \frac{d\theta_{-\bk_2}}{2\pi}e^{-i\theta_{-\bk_2}}\nonumber\\
&\times\!\!\prod_{\substack{\bku\\  \bk\neq \bk_1, \,\bk\neq -\bk_2}}\int \frac{d\theta_{\bk}}{2\pi}\,\,=0\, .
\end{align}
On the other hand, if $\bk_1=-\bk_2$ then $\langle \epsilon_{\bk_1}\epsilon_{\bk_2}\rangle_G=1$.
The calculation is analogous for $\bk_2 \in \mathrm{uhs}$ and $\bk_1 \in {\rm lhs}$. For $\bk_1$ and $\bk_2$ both in the upper hemisphere or both in the lower hemisphere, the correlation function aways vanishes, since $\bk_1\neq-\bk_2$. 

The non-Gaussian part of the probability distribution function does not contribute to the two-point correlation. To see this, let us look at the term in the second line of eq.~\eqref{Probability}. It gives a contribution of the form
\begin{align}
&\sum_{\bp\neq \bq\in {\rm uhs}}b(p,q, |\bp+\bq|)\int \frac{d\theta_{\bk_1}}{2\pi}e^{i\theta_{\bk_1}}
\int \frac{d\theta_{-\bk_2}}{2\pi}e^{-i\theta_{-\bk_2}}\nonumber\\
&\times\prod_{\substack{\bku\\  \bk\neq \bk_1, \,\bk\neq -\bk_2}}\int \frac{d\theta_{\bk}}{2\pi}
\cos(\theta_{\bp}+\theta_{\bq}-\theta_{\bp+\bq})\nonumber\, .
\end{align}
As for the one-point function, this integral always vanishes except in the case where $\bk_1=0$ or $\bk_2=0$, which corresponds to a background mode. A similar argument applies to the other terms of eq.~\eqref{Probability}. Consequently, the two-point correlation function is simply given by its Gaussian part. In the continuum limit it reads
\begin{equation}
\label{eps2}
\langle \epsilon(\bk_1)\epsilon(\bk_2)\rangle  =  \frac{(2\pi)^3}{V}\delta_D(\bk_1 + \bk_2)\, .
\end{equation}

This shows that there is no information in the two-point phase correlation either. This result is again a direct consequence of statistical homogeneity and isotropy, which enforce the two-point correlation function to be non-zero only for $\bk_1=-\bk_2$, in which case
$\epsilon(\bk_1)\epsilon(-\bk_1)=1$.

The lowest non-trivial phase correlation function is therefore the three-point correlation function. 

\subsection{The three-point correlation function}

To calculate the three-point correlation function, let us first consider the case where $\bk_1, \bk_2 \in \mathrm{uhs}$ and $\bk_3 \in \mathrm{lhs}$. The Gaussian term in the first line of eq.~\eqref{Probability} yields no contribution (except if one of the $\bk_i=0$ and the other two are opposite):
\begin{equation}
\prod_{\bku}\int  \frac{d\theta_{\bk}}{2\pi} e^{i\theta_{\bk_1}} e^{i\theta_{\bk_2}}e^{-i\theta_{-\bk_3}} = 0\, ,
\end{equation}
illustrating that the three-point phase correlation is zero for a Gaussian field. 

The non-Gaussian part on the other hand gives a non-zero contribution. As an example, let us look at the term in the second line of eq.~\eqref{Probability}. It gives a contribution of the form
\begin{align}
&\sum_{\bp\neq \bq\in {\rm uhs}}b(p,q, |\bp+\bq|)\int \frac{d\theta_{\bk_1}}{2\pi}e^{i\theta_{\bk_1}}
\int\frac{d\theta_{\bk_2}}{2\pi}e^{i\theta_{\bk_2}}\nonumber\\
&\int \frac{d\theta_{-\bk_3}}{2\pi}e^{-i\theta_{-\bk_3}}
\times\prod_{\substack{\bku\\  \bk\neq \bk_1, \bk_2\\ \bk\neq -\bk_3}}\int \frac{d\theta_{\bk}}{2\pi}
\cos(\theta_{\bp}+\theta_{\bq}-\theta_{\bp+\bq})\nonumber\, .
\end{align}
The integrals are non-zero when $\bq=\bk_1$, $\bp=\bk_2$ and $\bq+\bp=-\bk_3$ or when $\bq=\bk_2$, $\bp=\bk_1$ and $\bq+\bp=-\bk_3$ (recall that $\bq=-\bk_1$ or $\bq=-\bk_2$ are not possible here since $\bq, \bk_1$ and $\bk_2$ are all in the uhs). Combining these two cases, the expression above simply reduces to $b(k_1,k_2,-|\bk_1+\bk_2|)$. The other terms in eq.~\eqref{Probability} can be treated in the same way and the full result is
\begin{align}
&\langle \epsilon_{\bk_1}\epsilon_{\bk_2}\epsilon_{\bk_3}\rangle=\delta^K_{\bk_1+\bk_2+\bk_3}
\Bigg\{\left(\frac{\sqrt{\pi}}{2}\right)^3b(k_1,k_2,k_3)\nonumber\\
&+ \frac{\sqrt{\pi}}{12}b(k_1,k_2, k_3)\Big[\delta^K_{\bk_2-\bk_1}\!\!+\delta^K_{\bk_3-\bk_2}\!\!+\delta^K_{\bk_1-\bk_3} \Big] \Bigg\}\, .
\label{eps3_dis}
\end{align}
One can check that this result holds for all configurations of $\bk_1,\bk_2$ and $\bk_3$.

In the continuum limit, $V \rightarrow \infty$, the Kronecker delta becomes a Dirac delta function, $\delta_K \rightarrow \frac{(2\pi)^3}{V}\delta_D$. The second term has an extra factor of $V^{-1}$ over the first term causing it to be suppressed in this limit and become infinitesimal with respect to the first term. The three-point function therefore becomes
\begin{align}
\label{eps3}
\langle \epsilon(\bk_1)&\epsilon(\bk_2)\epsilon(\bk_3)\rangle=\\
&\frac{(2\pi)^3}{V}\left(\frac{\sqrt{\pi}}{2}\right)^3b(k_1,k_2,k_3)\,\delta_D(\bk_1+\bk_2+\bk_3)\, .\nonumber
\end{align}
This expression for the three-point function is very simple. It shows that at lowest order in perturbation theory the expectation value of the ratio becomes proportional to
\be
\left\langle\frac{\delta(\bk_1)\delta(\bk_2)\delta(\bk_3)}{|\delta(\bk_1)\delta(\bk_2)\delta(\bk_3)|} \right\rangle
\propto \frac{\langle \delta(\bk_1)\delta(\bk_2)\delta(\bk_3)\rangle}
{\sqrt{\langle|\delta(\bk_1)|^2 \rangle\langle|\delta(\bk_2)|^2 \rangle\langle|\delta(\bk_3)|^2 \rangle}}\, .
\ee
Note that this is not equivalent to the second line of eq.~\eqref{ratio}. This result follows directly from the form of the probability distribution function of the phase at lowest order in the Edgeworth expansion~\eqref{Probability}. At next order in perturbation theory, the probability distribution function depends on higher-order cumulants, adding new contributions to eq.~\eqref{eps3}. We will discuss in more detail these higher-order contributions in section~\ref{subsec:cumulants}.

\subsection{Correlations at second-order in perturbation theory}

Let us now calculate eq.~(22) explicitly by expressing the term $b(k_1,k_2,k_3)$ at lowest order in PT. The density contrast at second-order in PT, $\delta^{(2)}$, can be written as~\citep{Goroff1986a, Bouchet1995a, Bernardeau:2001qr}
\begin{align}
\delta^{(2)}(\bk, z)=&\int d^3\bq_1 \int d^3\bq_2\,\delta_D(\bk-\bq_1-\bq_2)\nonumber\\
&F_2(\bq_1,\bq_2, z) \delta^{(1)}(\bq_1, z)\delta^{(1)}(\bq_2, z)\, ,
\end{align}
where $\delta^{(1)}$ is the linear density contrast and the kernel $F_2$ is, for an arbitrary cosmology,
\bie
 F_2(\bq_1,\bq_2, z) &=& \frac{1}{2}(1+\epsilon) \, + \, \frac{\hat{\bq}_1 \cdot \hat{\bq}_2}{2} \left( \frac{q_1}{q_2}  + \, \frac{q_2}{q_1} \right)\nonumber\\
&& +\, \frac{1}{2}(1-\epsilon)(\hat{\bq}_1 \cdot \hat{\bq}_2)^2, 
\label{f2}
\eie
with $\epsilon(z) = \frac{3}{7} \left( \frac{\rho_m(z)}{\rho_{tot}(z)} \right)^{-\frac{1}{143}}$. Thus in $\Lambda$CDM cosmology, $\epsilon=\frac{3}{7}$ in excellent approximation ($\rm error\lesssim1\%$), from the beginning of the matter-dominated era (redshift $z\approx3600$) to today's mildly dark energy dominated era ($z=0$). The bispectrum at tree-level in PT is then
\be
\label{bisp}
B(k_1,k_2,k_3)=F_2(\bk_1,\bk_2, z)P_L(k_1, z)P_L(k_2, z) + {\rm cyc,}
\ee
where $P_L(k, z)$ is the linear power spectrum. Inserting this expression into eq.~\eqref{eps3}, we find for the three-point phase correlation function at second-order in PT
\begin{align}
\langle& \epsilon(\bk_1) \epsilon(\bk_2)\epsilon(\bk_3)\rangle\! =\! \left(\frac{\sqrt{\pi}}{2}\right)^3\!\! 
\left(\frac{(2\pi)^3}{V}\right)^{\!\!3/2} \!\!\!\!\delta_D(\bk_1 + \bk_2 + \bk_3)\nonumber  \\
& \times 2\left[F_2(\bk_1,\bk_2, z) \sqrt{\frac{P_L(k_1, z)P_L(k_2, z)}{P_L(k_3, z)}}+{\rm cyc.} \right] \, .
\label{PhaseCorrelation2}
\end{align}
 
Eq.~\eqref{PhaseCorrelation2} shows how the emergence of non-linearities in the density field generates three-point correlations between the phases. These correlations respect the statistical homogeneity and isotropy of the density field, since they exist only between modes related by $\bk_1+\bk_2+\bk_3=0$. The three-point phase correlation is a robust statistical measure, in the sense that its expectation value does not vanish. This is in contrast with the one-point phase correlation and the two-point phase correlation, which, even if affected by non-linearities~\citep{Ryden:1991fi, Soda:1991jz, Jain:1998it} contain no information {\it on average} (recall that the one-point function has a vanishing expectation value (see eq.~\eqref{eps1}) and the two-point function has a constant expectation value, insensitive to the density field (see eq.~\eqref{eps2})).
Eq.~\eqref{PhaseCorrelation2} shows that measuring the three-point phase correlations directly provides information about the non-linear gravitational evolution of the density field, i.e. about the non-linear kernel $F_2$.  
 
Interestingly, eq.~\eqref{PhaseCorrelation2} also demonstrates that the phase correlations are sensitive to the primordial amplitude $A$ of the density fluctuations, through $\sqrt{A}$. This may seem counter-intuitive as the phase is itself independent of the primordial amplitude of the density field.  However, the emergence of phase correlation is due to the non-linear growth of density perturbations, which depends strongly on the primordial amplitude (or similarly on $\sigma_8$). This also suggests that the modulus and the phase of the density contrast are not completely statistically independent.  

\section{The line correlation function}
\label{sec:line}

We can now calculate the line correlation function by inserting expression~\eqref{PhaseCorrelation2} into eq.~\eqref{ldef}. We find
\begin{align}
\ell(r, z)=&\frac{r^{9/2}}{8\sqrt{2}\,(2\pi)^3 }\hspace{-0.1cm}\underset{\substack{|\bk_1|, |\bk_2|,\\ |\bk_1 + \bk_2| \leq 2\pi/r}}{\int\!\!\int}
\!\!\!\!\!\!d^3\bk_1\,d^3\bk_2 \, F_2(\bk_1,\bk_2, z)\nonumber\\
&\times\sqrt{\frac{P_L(k_1, z)P_L(k_2, z)}{P_L(|\bk_1+\bk_2|, z)}}\,\Big[\cos\big((\bk_2-\bk_1)\cdot\br \big)\nonumber\\
&+2\cos\big((\bk_1+2\bk_2)\cdot\br \big) \Big]\, .
\end{align}
The integrand depends only on the length of the two wavevectors and on the angle between them, $\mu=\hat\bk_1\cdot\hat\bk_2$, so the integrals can be simplified to 
\begin{align}
\label{Line Correlation}
\ell(r, z) &= \frac{r^{9/2}}{8\pi\sqrt{2}}\int^{\frac{2\pi}{r}}_{0}\!dk_1\,k_1^2 \int^{\frac{2\pi}{r}}_{0}\!
dk_2\,k_2^2 \int^{\mu_{\rm cut}}_{-1} \! d\mu \\
& \times F_2(k_1,k_2,\mu, z)  \:  \sqrt{\frac{P_L(k_1, z)P_L(k_2, z)}{P_L(|\bk_1+\bk_2|, z)}}\nonumber\\
&\times \Big[j_0(|\bk_2-\bk_1|\cdot r)  + 2j_0(\bk_1+2\bk_2|\cdot r) \Big]\, ,\nonumber
\end{align}
with $j_0(x)\equiv\sin(x)/x$. The upper limit $\mu_{\rm cut}=\min\{1,\max\{-1,[(2\pi/r)^2-k_1^2-k_2^2]/[2k_1k_2])\}\}$ enforces the condition $|\bk_1 + \bk_2| \leq 2\pi/r$. 

Eq.~\eqref{Line Correlation} is the main result of this paper. It provides an analytical expression for the line correlation function, valid up to second-order in PT. As for the three-point correlation function in Fourier space, the line correlation function is a robust statistical measure, which probes the mean correlations of the phase along a line in real space. The average in eq.~\eqref{ldef} filters out the uncorrelated part from the individual shifts of the phase and extracts the correlated part, which is directly sensitive to the non-linear kernel $F_2$. 

In the regime where the bias between the matter density field and the galaxy density field can be described by a linear mapping $\delta_{\rm g}(\bk)=b_1\cdot \delta_{\rm dm}(\bk)$, the phase of the galaxy density contrast is the same as the phase of the dark matter density contrast
\be
\label{bias}
\eps_{\rm g}(\bk)=\frac{\delta_{\rm g}(\bk)}{|\delta_{\rm g}(\bk)|}=\frac{\delta_{\rm dm}(\bk)}{|\delta_{\rm dm}(\bk)|}=\eps_{\rm dm}(\bk)\, .
\ee
In this regime, expression~\eqref{Line Correlation} does therefore describe the phase correlations of the galaxy fluctuations as well as those of the dark matter fluctuations. Since eq.~\eqref{Line Correlation} is sensitive to the primordial amplitude of the density field, the line correlation function provides a way to measure $\sigma_8$ directly from galaxy clustering, without degeneracy with the linear bias parameter $b_1$. In the case of a non-linear or non-local bias however, the line correlation of galaxies is expected to receive corrections with respect to the line correlation of the dark matter. 

In a forthcoming paper we will study the contribution of the non-linear bias to the galaxy line correlation function. This will allow us to determine if the resilience of $\ell(r, z)$ to the linear bias $b_1$ can break the degeneracy between the primordial amplitude and $b_1$ and consequently improve measurements of $\sigma_8$. We will then investigate how well various cosmological parameters can be measured with the line correlation function and what improvements this new observable brings compared to standard bispectrum measurements. Note that since the two-point function of the phase is trivial (see eq.~\eqref{eps2}), the Gaussian contribution to the variance of $\ell(r)$ is independent of the power spectrum. In other words, the variance of $|\delta|$ does not contribute to the variance of the line correlation. This differs from the bispectrum which is sensitive to both the modulus and the phase of $\delta$ and whose variance is by consequence dominated by the variance of $|\delta|$. The line correlation function seems therefore better adapted to measure information beyond the linear regime, as it targets specifically the non-Gaussian part of the density field.

\section{Results}
\label{sec:num}

\subsection{The line correlation in a $\Lambda$CDM cosmology}
\label{sec:line_lin}
\begin{figure}[t]
\centering
\includegraphics[width=0.49\textwidth]{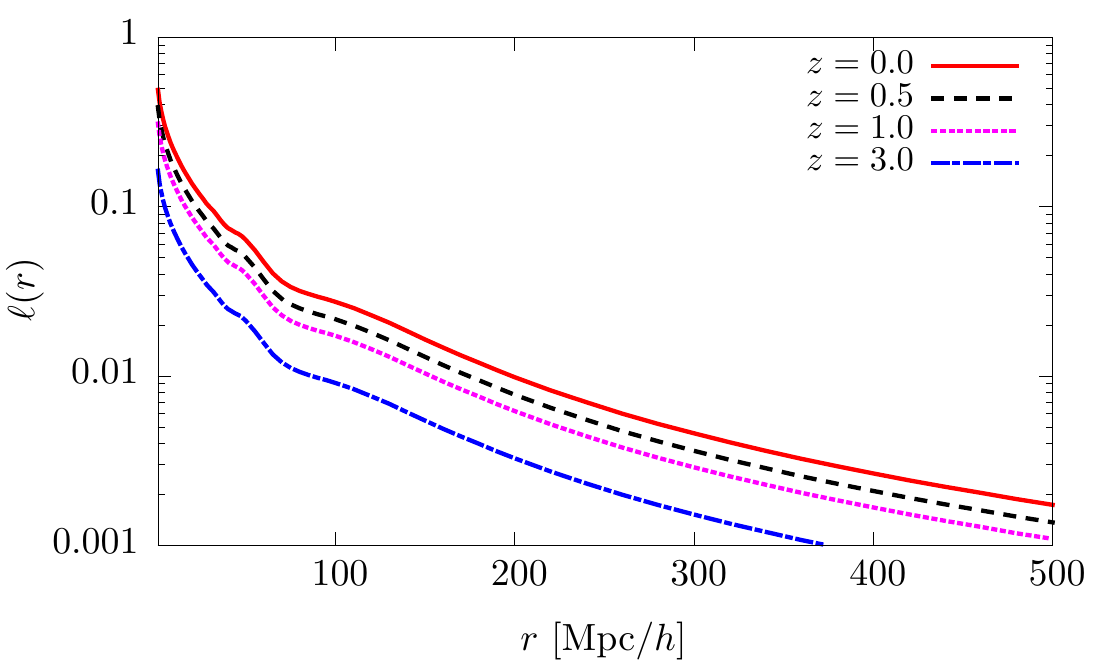}
\caption{The line correlation $\ell(r, z)$, calculated from eq.~\eqref{Line Correlation}, plotted as a function of $r$ at four different redshifts $z$. \label{fig:lineF2}}
\end{figure}

We calculate the line correlation function $\ell(r, z)$ in a $\Lambda$CDM universe with $\Omega_m = 0.25,\, \Omega_{\Lambda} = 0.75,\, h = 0.73$ (defined via the Hubble parameter $H_0=100\,h\,\rm km\,s^{-1}\,Mpc^{-1}$), $n_s=1$ and $\sigma_8 = 0.9$. We use \textsc{camb}~\citep{Lewis2000} to calculate the linear power spectrum. Our C code to calculate $\ell(r, z)$ can be downloaded at 
\href{http://www.blue-shift.ch/phase}{www.blue-shift.ch/phase}.

Figure~\ref{fig:lineF2} shows the line correlation as a function of the separation $r$ for four different redshifts. The line correlation function increases as the redshift decreases, reflecting how the non-linear growth of the density field enhances correlations between the phases. The growth of the line correlation with time is however slower than the growth of the n-point functions of the density field. The bispectrum for example grows as $D^4_1(a)$ ($D_1$ here denotes the linear growth rate of $\delta(\bk, z)$), whereas the line correlation function grows linearly only: $\ell(r,z)\propto D_1(a)$.

We see that the line correlation function decreases with the separation $r$. This is not surprising as non-linearities are more important at small scales. However, non-negligible correlations subsist up to relatively large separations. Looking at eq.~\eqref{Line Correlation}, we see that at large $k$ the integrand scales as 
\be
k^4\sqrt{P_L(k)} j_0(kr)\sim \frac{k^{3/2}}{r}\, ,
\ee
where we have used that $P_L(k)\sim k^{-3}$ and $j_0(kr)\sim (kr)^{-1}$ at large $k$. The integral is therefore dominated by the cut-off scale $k_{\rm max}=2\pi/r$ leading to 
\be
\ell(r, z)\sim r^{9/2} \times \frac{k_{\rm max}^{7/2}}{r}\sim r^0\, .
\ee
Due to the pre-factor $r^{9/2}$ the line correlation function is explicitly independent of $r$. However, since the integral is dominated by the cut-off scale, $\ell(r, z)$ probes the kernel $F_2$ at that scale, i.e. at $k=2\pi/r$. 
The line correlation is therefore truly a measurement of the correlations between the phases at the scale $r$.

Finally let us point out that the Baryon Acoustic Oscillations (BAO) are clearly visible at all redshifts in figure~\ref{fig:lineF2}. These oscillations reflect the impact of baryons in the emergence of correlations between the phase of the density field in the non-linear regime. In a forthcoming paper we will investigate if by combining measurements of the two-point correlation function and of the line correlation we can improve the determination of the BAO peak's position. 

\begin{figure}[!t]
\centerline{\includegraphics[width=0.49\textwidth]{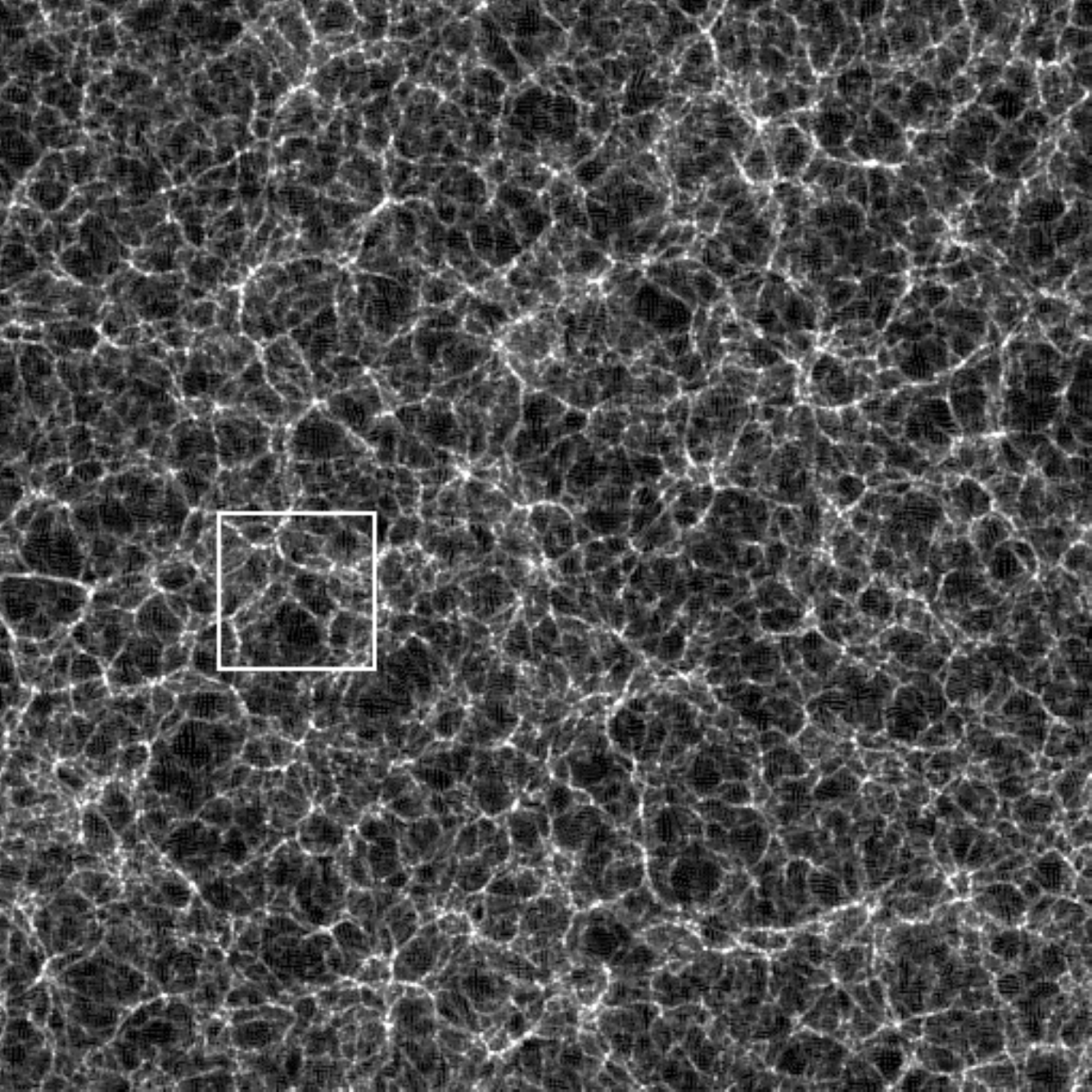}}
\centerline{\includegraphics[width=0.49\textwidth]{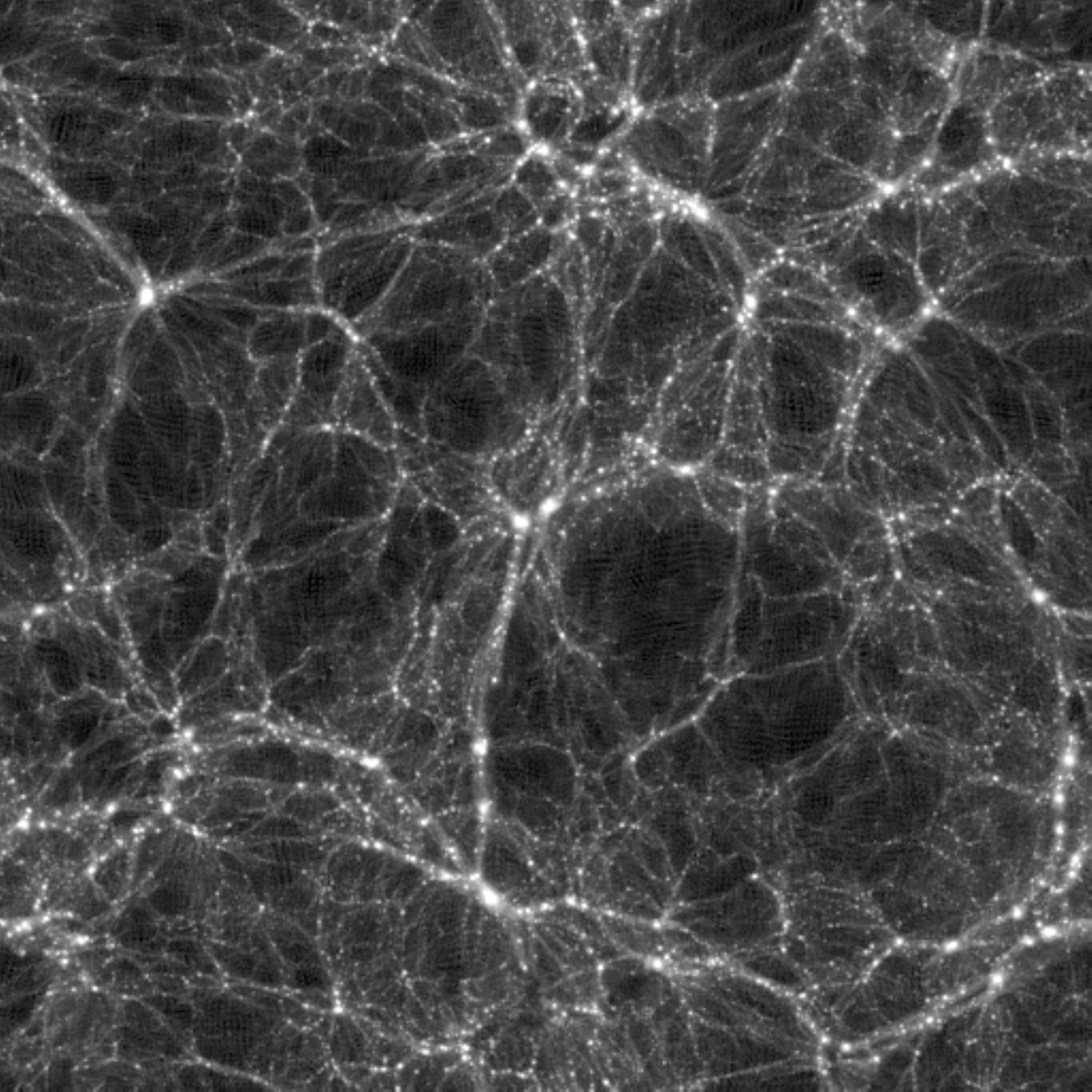}}
\caption{The top panel shows a two-dimensional projection of a $10 \,h^{-1} \, \mathrm{Mpc}$ thick tranche of one of a $1 \, h^{-1} \, \mathrm{Gpc}$ test simulation box ($512^3$ particles) at $z = 0$. The white box shows a scale comparison of the lower panel: a $10 \,h^{-1} \, \mathrm{Mpc}$ thick tranche of one of the $150 \, h^{-1} \, \mathrm{Mpc}$ simulation boxes at $z = 0$} 
\label{fig:sim}
\end{figure}
\subsection{Comparison with numerical simulations}

We now compare our analytical expression for the line correlation function~\eqref{Line Correlation} with direct measurements from a set of numerical N-body simulations using the \textsc{gadget}-2 code~\citep{Springel:2000yr, Springel:2005mi}.
In order to calculate the line correlation function over a large range of scales, we ran two sets of dark matter only $\Lambda\mathrm{CDM}$ cosmological simulations with Millennium cosmological parameters (as in section~\ref{sec:line_lin}). Five simulations were run in cubic boxes of comoving side-length of $L = 150\, h^{-1}\, \mathrm{Mpc}$ with $512^3$ particles and a further eight with $L = 1\, h^{-1}\, \mathrm{Gpc}$ with $256^3$ particles. The initial conditions were generated at $z = 99$ by displacing the particles from their grid positions according to second-order Lagrangian PT with an initial power spectrum generated from \textsc{camb}~\citep{Lewis2000}. Figure~\ref{fig:sim} shows the density field obtained from two of these simulations.

We measure the line correlation function at scales $r=1-30\,h^{-1}{\rm Mpc}$ from the first set of simulations and at scales $r=8-250\,h^{-1}{\rm Mpc}$ from the second set of simulations. Explicitly, this measurement is performed by evaluating the discretized eq.~(D4c) of \citet{Obreschkow2013a}, which assumes cosmic structure to be ergodic such that ensemble averages $\langle\rangle$ can be substituted for spatial averages. In a forthcoming paper, we will present in detail the code we used to measure $\ell(r, z)$. 
\begin{figure*}[]
\centering
\includegraphics[width=0.49\textwidth]{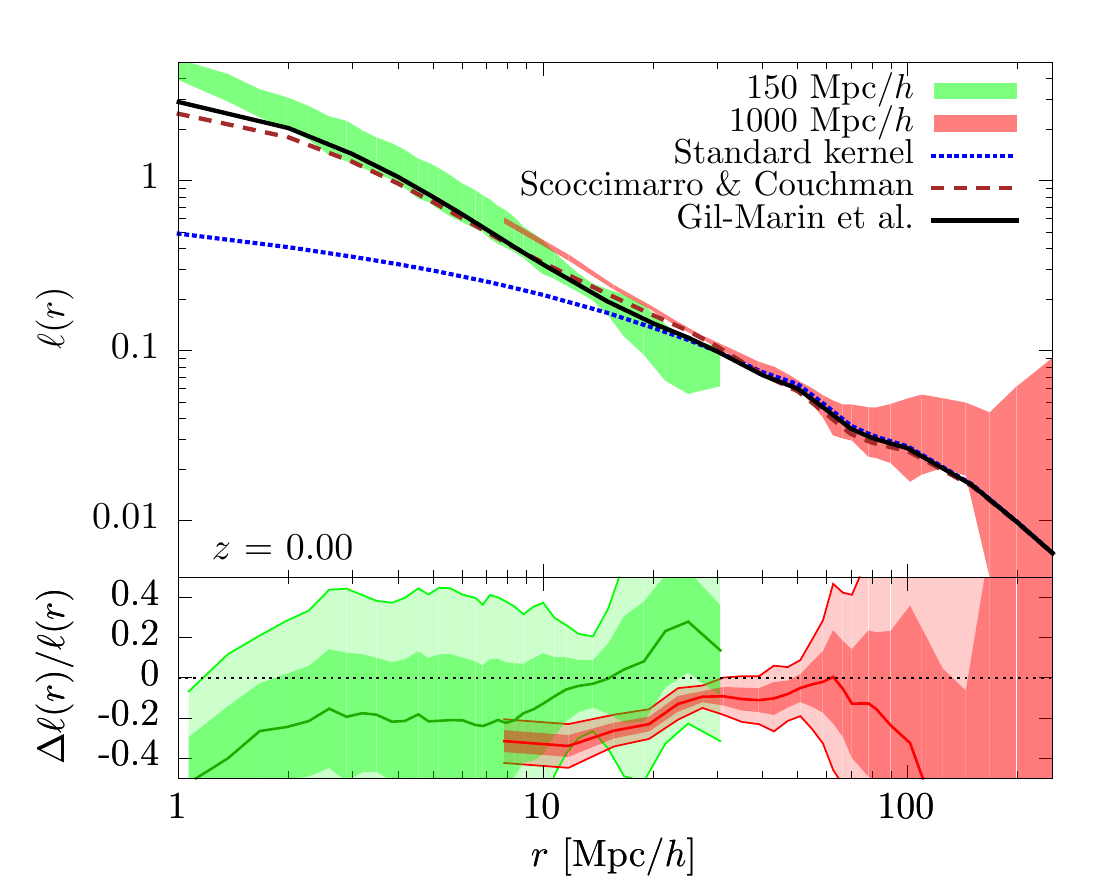}  \includegraphics[width=0.49\textwidth]{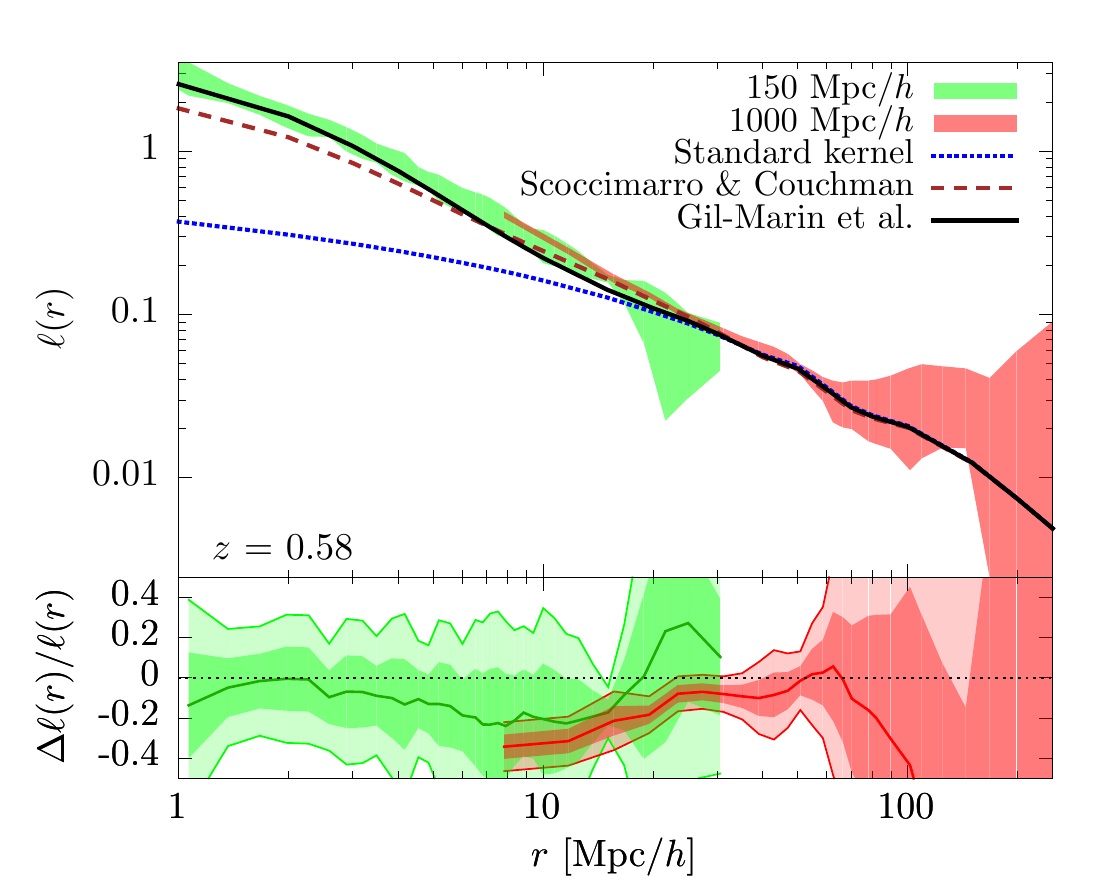}\\
\includegraphics[width=0.49\textwidth]{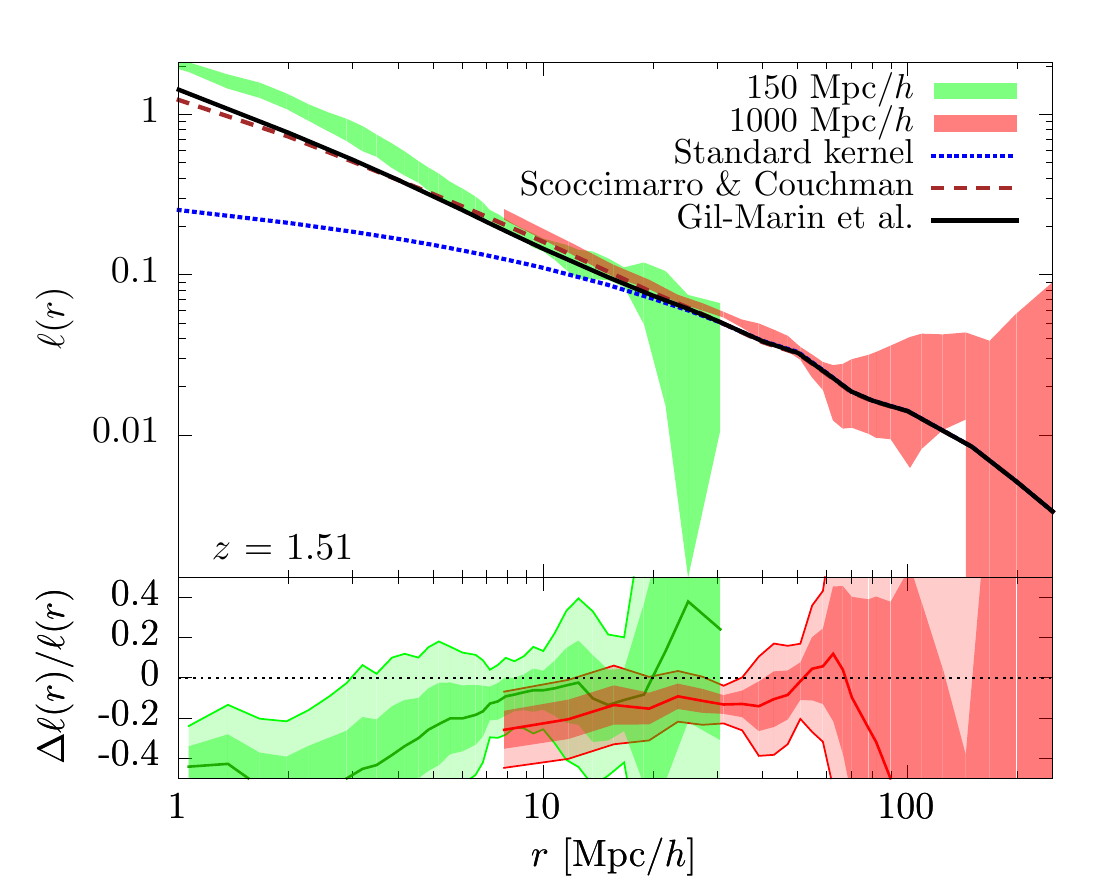}  \includegraphics[width=0.49\textwidth]{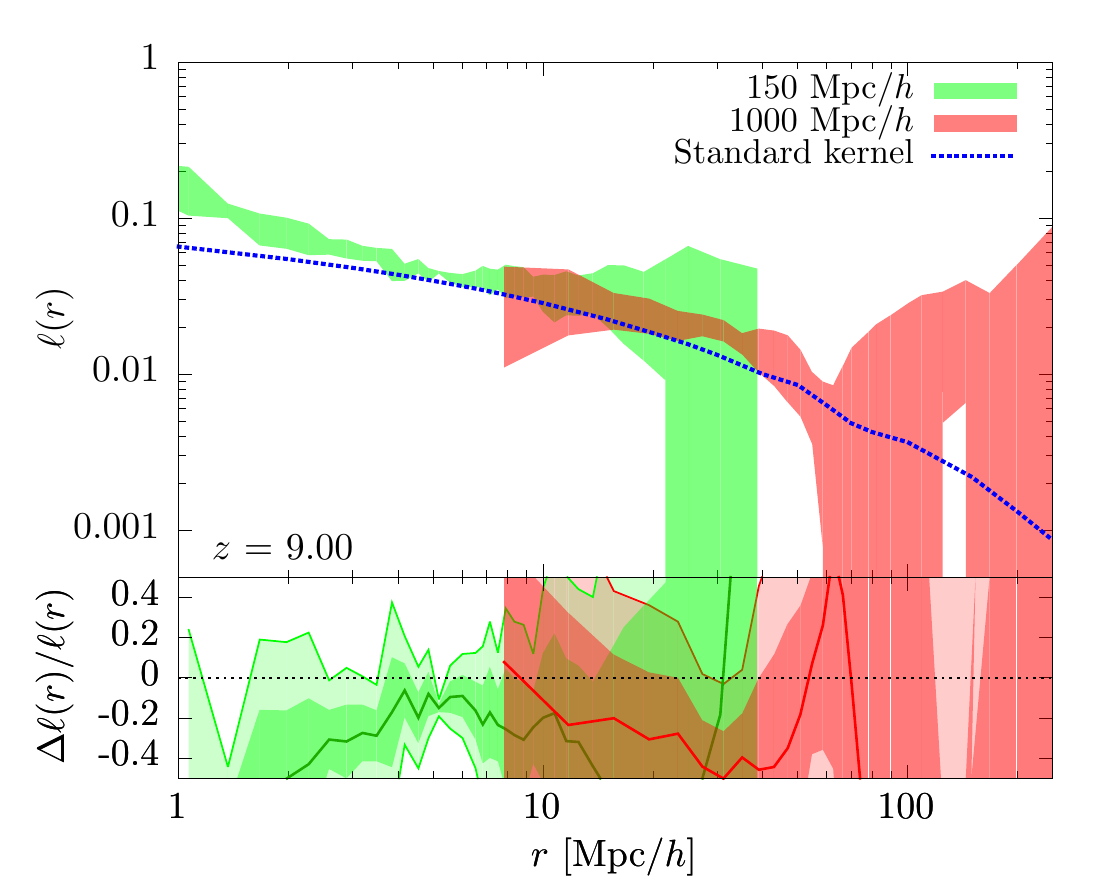}
\caption{Comparison of the analytical $\ell(r, z)$~\eqref{Line Correlation} (blue dotted line) with numerical simulations (red and green bands). The bands indicate the standard deviation across the simulations. The black solid line and the brown dashed line show the analytical result beyond second-order in PT, calculated with the fitted bispectrum of~\citet{Gil-Marin2012b} and~\citet{Scoccimarro2001b}, as discussed in section~\ref{sec:beyond_PT}. The lower panel of each subfigure shows the fractional difference between the theoretical curve, $\ell(r)$, and the mean value from the simulation results so that $\Delta \ell(r) = \ell(r) - \ell_{\mathrm{sim}}(r)$. For $z = 0.00, 0.58$ and $1.51$, $\ell(r)$ is calculated with the Gil-Mar\'{\i}n et al. kernel, whereas for $z=9.00$ the standard kernel is used. The darker and lighter shaded red and green regions show respectively the 1-$\sigma$ and 2-$\sigma$ deviations from the simulation means\label{redshiftcomp} which are themselves given by the central solid lines. The thinner solid lines simply border the 2-$\sigma$ regions to guide the eye. }
\end{figure*}
In figure~\ref{redshiftcomp} we compare the numerical results (red and green bands), measured from four different snapshots, with our expression~\eqref{Line Correlation} (blue dotted line). The numerical data are shown as shaded regions, representing the standard deviation of the different simulation runs. The line correlations measured in different simulation runs naturally differ due to so-called cosmic variance, i.e. the variance amongst independent realizations of the density field, truncated to a cubic box of $150\,h^{-1}{\rm Mpc}$ and $1\,h^{-1}{\rm Gpc}$ sides. A cubic flat-top window function with the size of the grid-cells was used for the numerical evaluation of $\ell(r)$, since this function minimizes the noise equivalent bandwidth. However, by testing different window functions, including the smoothing kernel of the N-body particles, the systematic errors induced by the window function turned out to be much smaller that the shot noise of the simulations.

At very high redshift (bottom panel, z=9) our analytical expression reproduces the numerical result within 2-$\sigma$ at almost all scales measured and within 1-$\sigma$ for $r \gsim 8 \, h^{-1} \mathrm{Mpc}$ up to where the larger simulation becomes volume limited. At lower redshifts, the analytical result agrees within 1-$\sigma$ with the numerical simulations above $r\gsim 30\,h^{-1}{\rm Mpc}$. Below $30\,h^{-1}{\rm Mpc}$ however, our expression underestimates the line correlation function by up to a factor 10 at $r\simeq1\,h^{-1}{\rm Mpc}$. To describe what happens in this regime, we need to go beyond the second order in PT. 

\subsection{Beyond second order in perturbation theory}
\label{sec:beyond_PT}

Equation~\eqref{Line Correlation} has been calculated at the lowest order in PT and is therefore a valid approximation only in the mildly non-linear regime. Once gravitational evolution becomes strongly non-linear, higher-order contributions have to be included.
From eqs.~\eqref{eps3} and~\eqref{defb} we see that two types of corrections in the three-point phase correlation become relevant in the non-linear regime: loop corrections to the power spectrum and loop corrections to the bispectrum. Schematically, at lowest order in PT, the three-point phase correlation scales as $\delta^{(1)}$, whereas the next-order corrections from the power spectrum and the bispectrum are both proportional to $\left(\delta^{(1)}\right)^3$. We expect therefore these two types of corrections to become relevant at the same scale.

In addition to these two corrections, we expect also corrections from the higher-order cumulants in the non-Gaussian probability distribution function~\eqref{prob_gen}, which have been neglected from eq.~\eqref{P_Edg} onwards. One can show that the fourth and higher-order even cumulants have exactly zero contribution to the three-point phase correlation whilst certain configurations of the higher-order odd cumulants do contribute. For example, the fifth cumulant, $\langle \delta(\bk_1) \delta(\bk_2)  \delta(\bk_3)  \delta(\bk_4)   \delta(\bk_5) \rangle_c$, contributes to the three-point phase correlation  when two of the $\bk_i$ are opposite. This term also induces corrections to the line correlation of the order $\left(\delta^{(1)}\right)^3$ and it is therefore potentially similarly important as the power spectrum and the bispectrum corrections. 

These three types of corrections can in principle be calculated within the framework of PT. They involve however a large number of configurations, and are consequently not trivial to implement. Moreover, this strategy would only allow us to extend the range of validity of our expression by a limited amount and it certainly would not take us into the strongly non-linear regime. 

We adopt therefore a different approach: we use numerical fits to the power spectrum and the non-linear coupling kernel to calculate $\ell(r, z)$ in the strongly non-linear regime. For the power spectrum we use the well-known halo-fit power spectrum~\citep{Smith:2002dz, 2012ApJ...761..152T}. For the non-linear coupling kernel we use an effective kernel which interpolates between the standard non-linear kernel $F_2$ in eq.~\eqref{f2} and numerical fits at small scales~\citep{Scoccimarro1999c, Scoccimarro2001b, Gil-Marin2012b}. This modification to the non-linear coupling kernel, referred to as hyperextended PT, uses a set of parameters fitted to numerical simulations to construct the bispectrum
\be
B(k_1,k_2,k_3)=F_2^{\mathrm{eff}}(\bk_1,\bk_2, z)P(k_1, z)P(k_2, z) + {\rm cyc,}
\ee
with $P(k, z)$ the non-linear power spectrum. We use two different fitted kernels to calculate the line correlation function: the kernel proposed by~\citet{Scoccimarro2001b}, and the kernel proposed by~\citet{Gil-Marin2012b}, 
which provides a better fit to N-body data in the region $0.03\,h\,\mathrm{Mpc}^{-1} \leq k \leq 0.4\,h\,\mathrm{Mpc}^{-1}$ and redshift range $0 \leq z \leq 1.5$. 
\begin{figure}[]
\centerline{\includegraphics[width=0.49\textwidth]{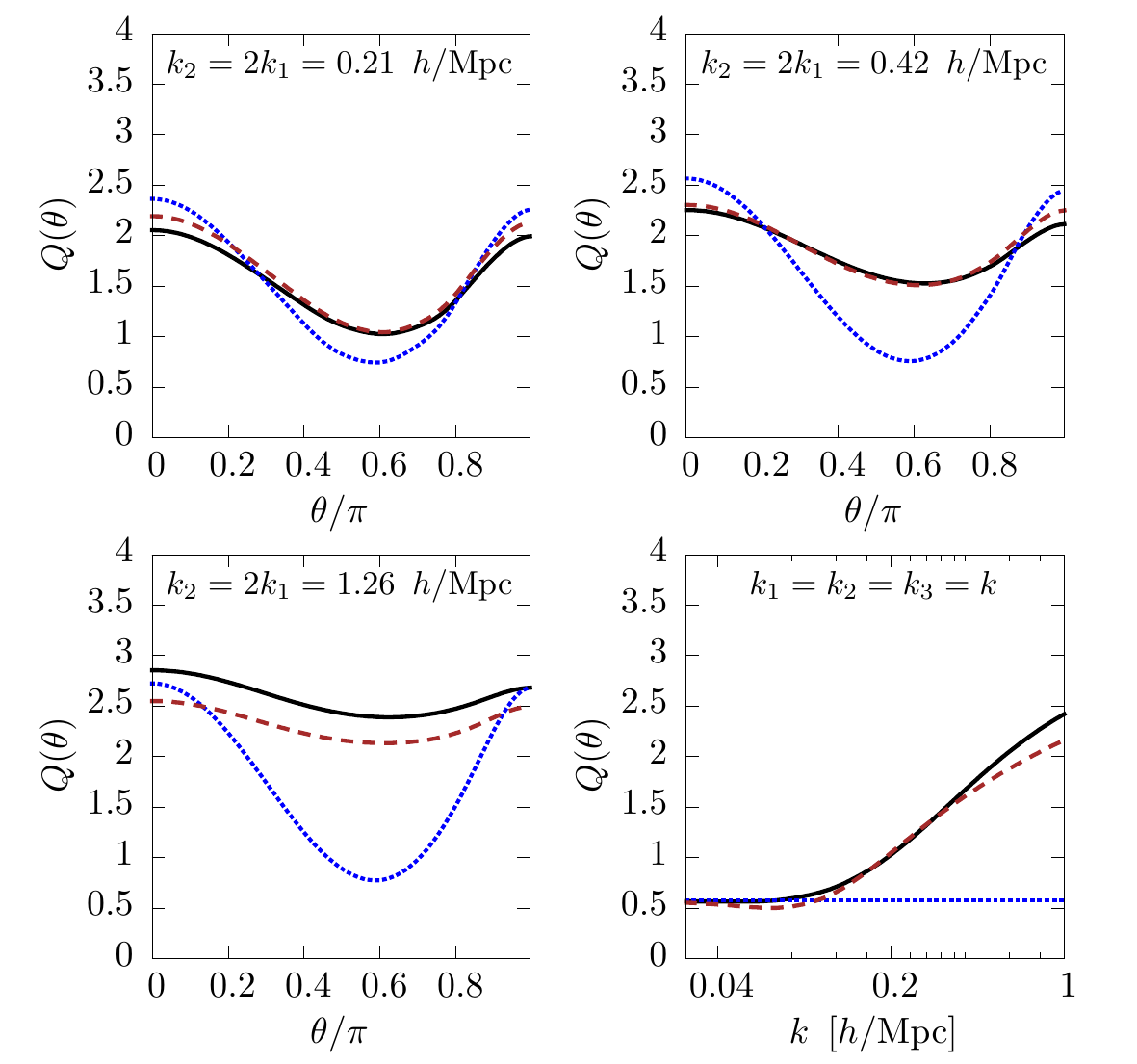}}
\caption{The upper left and right and lower left panels show the reduced bispectrum as a function of angle $\theta$ between $\bk_1$ and $\bk_2$ for two fixed scales, as shown in each panel. The blue dotted line shows the bispectrum described with the standard PT kernel of eq.~\eqref{f2} and the linear power spectrum, the brown dashed line uses the kernel of~\citet{Scoccimarro2001b} and the black solid line the kernel of~\citet{Gil-Marin2012b}, both with the non-linear power spectrum. The lower right panel shoes the reduced bispectrum for equilateral shapes.} 
\label{qangle}
\end{figure}

In figure~\ref{qangle} we plot the reduced bispectrum, defined by
\begin{equation}
Q = \frac{B(\bk_1,\bk_2,\bk_3)}{P(k_1)P(k_2) + P(k_2)P(k_3) + P(k_1)P(k_3)}\, ,
\label{qparam}
\end{equation}
for the different coupling kernels to illustrate the need for extensions to the analytical model beyond second-order PT. We see significant deviations between the non-linear effective kernels and the second-order PT. The analytical bispectrum calculated using hyperextended perturbation theory offers a more realistic fit to the numerical data, as shown in \citet{Scoccimarro2001b} and in~\citet{Gil-Marin2012b}. Note that since we use directly the parameters of the fitted kernels proposed in these two papers to calculate $\ell(r, z)$, any inaccuracy in the modelling of the bispectrum will propagate to the line correlation. Our goal is however not to find a perfect fit for the line correlation in the strongly non-linear regime, but rather to understand if non-linear effects in the bispectrum are responsible for the difference between the numerical result and our analytical expression at second order in PT.

In figure~\ref{redshiftcomp}, the brown dashed line shows the line correlation function calculated with the kernel of \citet{Scoccimarro2001b} and the black solid line shows the line correlation calculated with the kernel of \citet{Gil-Marin2012b}. The two kernels give very similar results and are in a 1-$\sigma$ agreement with the numerical results of the higher resolution simulations down to $r\simeq 2-3 \,h^{-1}{\rm Mpc}$~\footnote{Note that the parameters of the Gil-Mar{\'in} et al. kernel are only fitted up to $k \leq 0.4\,h\,\mathrm{Mpc}^{-1}$. For  $r < 15.7\,h^{-1}\,\mathrm{Mpc}$ the integral in eq.~\eqref{Line Correlation} involves larger $k$, outside the range of fitting. However even at these scales, our expression for the line correlation function seems to reproduce well the numerical results.}. We note however a tension between the results of the two sets of simulations in the range $10 < r < 30 \, h^{-1}\mathrm{Mpc}$, suggesting that the lower resolution simulations systematically overestimate the line correlation on these scales. Initially we tested $1\,h^{-1}\mathrm{Gpc}$ simulations with $128^3$ particles and for these there was an even greater excess of power with respect to the $512^3$ simulations and the analytical calculation.

For $r\gsim 20\,h^{-1}\,\mathrm{Mpc}$, the non-linear line correlation functions converge to the second-order perturbed expression~\eqref{Line Correlation} (blue dotted line), as expected from the fact that these scales only invoke modes $k\leq2\pi/r\approx0.3\,h\,\rm Mpc^{-1}$ of the linear regime in the definition (2) of $\ell(r,z)$.

Note that the method presented here only includes contributions from higher-order corrections in the power spectrum and the bispectrum. It does not take into account the third type of corrections due to the higher-order cumulants (above the 3rd moment) in the probability distribution function $\mathcal{P}[\delta]$. Calculating these higher-order cumulants is not trivial, as they contain a large number of terms. Below we propose a method to test the importance of these corrections directly from the numerical simulations.    

\subsection{Importance of higher-order cumulants}

\label{subsec:cumulants}

To evaluate the contribution of higher-order cumulants, neglected in the Edgeworth expansion eq.~\eqref{P_Edg}, let us reconsider the exact line correlation as defined in eq.~\eqref{ldef}. Following our discretization rules this equation becomes,
\be
\label{l_num_av}
\ell(r)=\left(\frac{r^3}{V}\right)^{\frac{3}{2}}\hspace{-0.6cm}
\sum_{\substack{|\bk_1|, |\bk_2|, \\ |\bk_1+\bk_2|\leq 2\pi/r}}\hspace{-0.3cm}\!\!\!\!j_0\big(|\bk_1-\bk_2|r\big)
\langle\epsilon_{\bk_1}\epsilon_{\bk_2}\epsilon_{-\bk_1-\bk_2}\rangle.
\ee
Neglecting the higher-order cumulants in passing from eq.~\eqref{prob_gen} to~\eqref{P_Edg} is equivalent to replacing $\langle\epsilon_{\bk_1}\epsilon_{\bk_2}\epsilon_{-\bk_1-\bk_2}\rangle$ by $(\sqrt{\pi}/2)^3\langle\delta_{\bk_1}\delta_{\bk_2}\delta_{-\bk_1-\bk_2}\rangle/\!\sqrt{\langle|\delta_{\bk_1}|^2\rangle\langle|\delta_{\bk_2}|^2\rangle\langle|\delta_{\bk_1+\bk_2}|^2\rangle}$, 
as can be seen from eqs.~\eqref{eps3_dis} and~\eqref{defb}. Thus, the discretized line correlation in the Edgeworth expansion reads
\be
\begin{split}
\label{hatl}
\hat\ell(r)=&\left(\frac{r^3}{V}\right)^{3/2}\left(\frac{\sqrt{\pi}}{2}\right)^3\hspace{-0.5cm}
\sum_{\substack{|\bk_1|, |\bk_2|, \\ |\bk_1+\bk_2|\leq 2\pi/r}}j_0\big(|\bk_1-\bk_2|r\big)\\
&\times\frac{\langle\delta_{\bk_1}\delta_{\bk_2}\delta_{-\bk_1-\bk_2}\rangle}
{\sqrt{\langle\delta_{\bk_1}\delta_{-\bk_1}\rangle\langle\delta_{\bk_2}\delta_{-\bk_2}\rangle
\langle\delta_{\bk_1+\bk_2}\delta_{-\bk_1-\bk_2}\rangle}}.
\end{split}
\ee
The contribution of the neglected higher-order cumulants is then given by the difference $\chi(r)\equiv \ell(r)-\hat\ell(r)$. To evaluate this difference numerically for a particular N-body simulation, we assume ergodicity to substitute ensemble averages $\langle\rangle$ for spatial averages over all translations and generalized rotations of a particular shape $(\bk_1,\bk_2,-\bk_1-\bk_2)$. In eq.~\eqref{l_num_av}, the averages $\langle\rangle$ can then simply be dropped, since the sums over $\bk_1$ and $\bk_2$ already incorporate a spatial average of all triangles of a fixed shape. In eq.~\eqref{hatl}, the same reasoning can be applied to drop the averages $\langle\rangle$ in the numerator $\langle\delta_{\bk_1}\delta_{\bk_2}\delta_{-\bk_1-\bk_2}\rangle$. In conclusion, the difference between eqs.~\eqref{l_num_av} and~\eqref{hatl} becomes
\begin{align}
\label{chi}
&\chi(r)=\left(\frac{r^3}{V}\right)^{3/2}\hspace{-0.5cm}
\sum_{\substack{|\bk_1|, |\bk_2|, \\ |\bk_1+\bk_2|\leq 2\pi/r}}j_0\big(|\bk_1-\bk_2|r\big)\,\times\\
&\left[\frac{\delta_{\bk_1}\delta_{\bk_2}\delta_{-\bk_1-\bk_2}}
{|\delta_{\bk_1}\delta_{\bk_2}\delta_{-\bk_1-\bk_2}|}-\left(\!\frac{\sqrt{\pi}}{2}\right)^3\!\!\!\!\frac{\delta_{\bk_1}\delta_{\bk_2}\delta_{-\bk_1-\bk_2}}
{\sqrt{P(k_1)P(k_2)P(|\bk_1+\bk_2|)}}\right].\nonumber
\end{align}
where $P(k)$ is the spatial average of $\delta_\bk \delta _{-\bk}$.

\begin{figure}[t]
\centering
\includegraphics[width=0.49\textwidth]{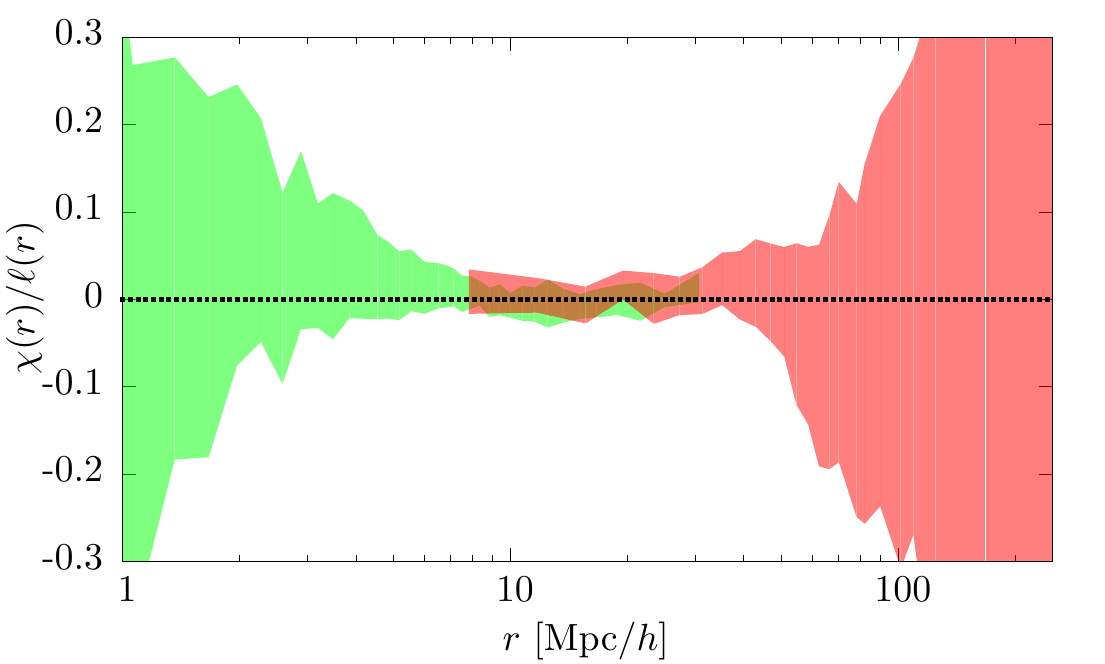}
\caption{The difference $\chi(r)=\ell(r)-\hat\ell(r)$ normalised by $\ell(r)$ calculated from our set of numerical simulations at $z = 0$ along with the 1-$\sigma$ deviation. We see that the function is compatible with zero albeit with large variances at both small and large scales. The latter is simply a result of cosmic variance.} \label{fig:l_diff}
\end{figure}

In figure~\ref{fig:l_diff}, we plot $\chi(r)$ for our set of numerical simulations, normalised by $\ell(r)$. We see that whilst this function is completely compatible with zero in the range of scales probed, there is a large variance at both small and large scales, i.e. $r \lsim 8 h^{-1}\mathrm{Mpc}$ and $r \gsim 40 h^{-1}\mathrm{Mpc}$. At large scales this is simply due to the cosmic variance of the simulations but at the smaller scales we need to either run more high resolution simulations or use other methods to determine the relevance of higher-order cumulants. In a forthcoming paper, we will calculate the line correlation function in the halo-model, to determine if, in the highly non-linear regime, the higher-order cumulants can be neglected as well.   

\section{Conclusion}
\label{sec:conclusion}

In this paper, we have derived an analytical expression for a novel estimator of cosmic large-scale structure: the so-called line correlation function. This function, introduced in~\citet{Obreschkow2013a}, is constructed from the three-point correlation of the phase of the density field. The line correlation is designed to target information beyond the power spectrum, in particular information relevant for the filamentary geometry of the cosmic web. 

We have shown that in the mildly non-linear regime, where perturbation theory is valid, the line correlation function can be expressed in terms of the non-linear kernel $F_2$. This reveals how the non-linear gravitational evolution of the density field generates correlations between phases that are initially randomly distributed and uncorrelated. We have compared our analytical expression with the results from N-body simulations. At $z=0$, the tree-level analytical result is in a 1-$\sigma$ agreement with the simulations for separations $r\gsim 30\,h^{-1}{\rm Mpc}$. At smaller separations however, our approximation using PT underestimates the line correlation function. We have used fitting formulae for the power spectrum and the non-linear coupling kernel to extend the validity of our expression into the strongly non-linear regime. With those, we found 1-$\sigma$ agreements with the numerical simulations down to $r\simeq 2-3 \,h^{-1}{\rm Mpc}$. 

Our analytical expression for the line correlation function has revealed two crucial advantages of this observable with respect to more standard statistical measures like the bispectrum of the density field. First, the line correlation function is by construction independent of the linear galaxy bias $b_1$, see eq.~\eqref{bias}. It has therefore the potential to break the degeneracy between measurement of $\sigma_8$ and the bias in galaxy surveys. Second, the line correlation function is independent of the modulus of the density field, whereas the bispectrum is not. As a consequence, the variance of the bispectrum is dominated by the variance of the modulus $|\delta|$, i.e. by the Gaussian part of the density field. The variance of the line correlation function on the other hand has a Gaussian part which is independent on the variance of $|\delta|$. It remains to be seen how this particularity of the line correlation can improve measurement of information
beyond the linear Gaussian regime.

In a forthcoming paper, we will use our analytical expression for the line correlation function to study in detail how well it can measure cosmological parameters, and which kind of degeneracy it can break with respect to power spectrum and bispectrum measurements. We will also extend our analytical expression to the more general situation of a non-linear bias parameter, which is expected to contribute to the phase correlations as well. Let us finally mention that our calculation has been performed in real space, and that we have consequently neglected any redshift-space distortions. These distortions can potentially contribute to the correlations between the phases and break the statistical isotropy of $\ell(r)$. In a future project, we will study the sensitivity of the line correlation to redshift-space distortions.

The C code used to calculate the line correlation function at lowest order in PT as well as with the Gil-Mar{\'in} fitted kernel can be downloaded at \href{http://www.blue-shift.ch/phase}{www.blue-shift.ch/phase}. 

\section*{Acknowledgements}

We thank Chris Power for running the high-resolution ($512^3$) simulations and for his great help in running the lower resolution simulations. It is also a pleasure to thank Mustafa Amin, Francis Bernardeau, Anthony Challinor, Enrique Gaztanaga, Eugene Lim and Patrick Valageas for useful and interesting discussions. We also thank the anonymous referee for constructive and useful comments. CB is supported by King's College Cambridge. Part of the research presented in this paper was undertaken
as part of the Survey Simulation Pipeline (SSimPL). DO and RW were supported by the Research Collaboration Award 12105012
of the University of Western Australia.

\bibliographystyle{apj}
\bibliography{library}

\end{document}